\documentclass[%
 superscriptaddress,
 amsmath,amssymb,
 aps,
 prl,
 reprint,
floatfix,
longbibliography
]{revtex4-2} 

\AtBeginDocument{%
	\newwrite\bibnotes
	\def\bibnotesext{Mn3Sn_paper_2.bib}
	\immediate\openout\bibnotes=\jobname\bibnotesext
	\immediate\write\bibnotes{@CONTROL{REVTEX42Control}}
	\immediate\write\bibnotes{@CONTROL{%
			apsrev42Control,author="08",editor="1",pages="1",title="0",year="1"}}
	\if@filesw
	\immediate\write\@auxout{\string\cite{apsrev42Control}}%
	\fi
	}
\usepackage{graphicx}
\usepackage{comment}
\usepackage{dcolumn}
\usepackage{bm}
\usepackage{lipsum}
\usepackage[colorlinks=true,linkcolor=blue,plainpages=false]{hyperref}
\usepackage[all]{hypcap} 
\usepackage{textcomp}

\newcommand{\subt}[1]{\textsubscript{#1}}

\begin{document}

\title{Time-dependent multistate switching of topological antiferromagnetic order in Mn$_3$Sn
}

\author{Gunasheel Kauwtilyaa Krishnaswamy}
 \affiliation{Department of Materials, ETH Zurich, 8093 Zurich, Switzerland}
\author{Giacomo Sala}
 \affiliation{Department of Materials, ETH Zurich, 8093 Zurich, Switzerland}
\author{Benjamin Jacot}
\affiliation{Department of Materials, ETH Zurich, 8093 Zurich, Switzerland}
\author{Charles-Henri Lambert}
\affiliation{Department of Materials, ETH Zurich, 8093 Zurich, Switzerland}
\author{Richard Schlitz}
\affiliation{Department of Materials, ETH Zurich, 8093 Zurich, Switzerland}
\author{Marta D. Rossell}
\affiliation{Electron Microscopy Center, Empa, Swiss Federal Laboratories for Material Science and Technology, Dübendorf, Switzerland}
\author{Paul Nöel}
\affiliation{Department of Materials, ETH Zurich, 8093 Zurich, Switzerland}
\author{Pietro Gambardella}
\affiliation{Department of Materials, ETH Zurich, 8093 Zurich, Switzerland}

\date{\today}

\begin{abstract}
The manipulation of antiferromagnetic order by means of spin-orbit torques opens unprecedented opportunities to exploit the dynamics of antiferromagnets in spintronic devices. In this work, we investigate the current-induced switching of the magnetic octupole vector in the Weyl antiferromagnet Mn$_3$Sn as a function of pulse shape, magnetic field, temperature, and time. We find that the switching behavior can be either bistable or tristable depending on the temporal structure of the current pulses. Time-resolved Hall effect measurements performed during the current pulsing reveal that Mn$_3$Sn switching proceeds via a two-step demagnetization-remagnetization process caused by self-heating over a timescale of tens of ns followed by cooling in the presence of spin-orbit torques. Single-shot switching measurements with 50~ps temporal resolution indicate that chiral spin rotation is either damped or incoherent in polycrystalline Mn$_3$Sn. Our results shed light on the switching dynamics of Mn$_3$Sn and prove the existence of extrinsic limits on its switching speed.

\end{abstract}
\maketitle
\section{Introduction}
Electric control of magnetic order in antiferromagnets has raised prospects for realizing high-speed and high-density magnetoelectric devices using materials with zero net magnetization \cite{Wadley2016,Olejnik2017,Meinert2018,DuttaGupta2020,Siddiqui2020,Arpaci2021}. The switching of the order parameter in antiferromagnets is achieved by either injecting spin currents from an adjacent heavy metal layer or current-induced spin-orbit torques intrinsic to noncentrosymmetric crystals \cite{Manchon2019}. Electrical readout, however, is problematic because of the small magnetoresistance \cite{Marti2014}, resistive artefacts \cite{Chiang2019,Jacot2020, Matalla-Wagner2020}, and absence of Hall effect in most conventional antiferromagnets. This problem can be elegantly solved by turning to noncollinear antiferromagnets, which combine topologically nontrivial electronic properties with chiral magnetic order. In these systems, the broken time-reversal symmetry and large Berry curvature in momentum space give rise to strong anomalous Hall effect (AHE) \cite{Nakatsuji2015,Nayak2016} and magneto-optical responses \cite{Higo2018b,Balk2019,Zhao2021a,Uchimura2022}, similar to ferromagnets but in the absence of significant magnetization. Theoretical work shows that these materials can even exhibit a large tunneling magnetoresistance \cite{Dong2021}, whereas the emergence of exotic phenomena such as the chiral anomaly \cite{Kuroda2017} and magnetic spin Hall effect \cite{Kimata2019,Kondou2021, Hu2021, Ghosh2022} makes them a very interesting playground for investigating the interplay of topology, electron transport, and magnetism \cite{Gomonay2014,Baltz2018, Smejkal2018}.

\begin{figure}[!htbp]
	\includegraphics[width=1\columnwidth,height=\textheight,keepaspectratio]{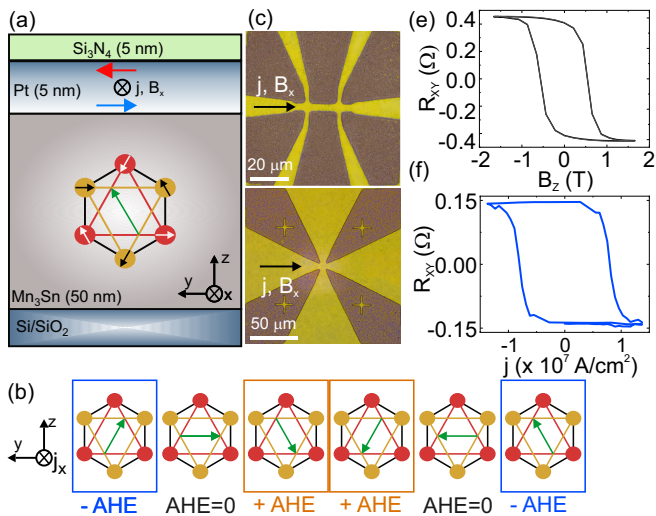}\
	\label{fig:1}
	\caption{(a) Cross-section of the Mn$_3$Sn/Pt bilayer. The inverted triangular spin structure is shown in the center: white and black arrows represent the Mn spins and the green arrow the octupole vector \textbf{g}. (b) Possible orientations of \textbf{g} and corresponding AHE signal. (c) Microscope image of a Hall bar device and (d) Hall cross used for switching and time-resolved measurements. (e) AHE of Mn$_3$Sn/Pt as a function of magnetic field along $z$ and (f) current density for 10~$\mu$s-long pulses and $B_{\rm x}=+200$~mT.}
\end{figure}

A prime candidate of this material class is Mn$_3$Sn, a hexagonal Weyl metal in which the Mn atoms form kagome lattice planes stacked along the $c$-axis with an inverse-triangular spin structure and all the spins oriented in-plane \cite{Tomiyoshi1982,Ohmori1987,Duan2015,Nakatsuji2015,Song2020}. The non-collinear antiferromagnetic order is best described by the magnetic octupole moment \textbf{g} of the six Mn spins that reside in two stacked inverted triangles on adjacent kagome layers [green arrow in Fig.~\ref{fig:1} (a)]. Magnetic anisotropy defines six possible orientations of the \textbf{g}-vector in the kagome plane [Fig.~\ref{fig:1} (b)]. The almost perfect 120$^{\circ}$ non-collinear spin alignment is slightly distorted by magnetic anisotropy, which leads to a weak ferromagnetic moment of $\sim0.002$~$\mu_B$ per Mn atom in the direction of the \textbf{g}-vector. This conveniently allows for the manipulation of antiferromagnetic order by external magnetic fields, whereas the large AHE and anomalous Nernst effect (ANE) of Mn$_3$Sn provide direct information on the orientation of \textbf{g} \cite{Ikhlas2017,Li2017,Ikeda2020,Nakano2021}. Importantly for applications, the topological properties of Mn$_3$Sn emerge in both polycrystalline and epitaxial thin films \cite{Higo2018a, Ikeda2019,You2019, Ikeda2020, Yoon2020, Yoon2021, Nakano2021, Khadka2020}.

\setlength{\textfloatsep}{10pt} 

\begin{figure*}[t]
	\includegraphics[width=\textwidth,keepaspectratio]{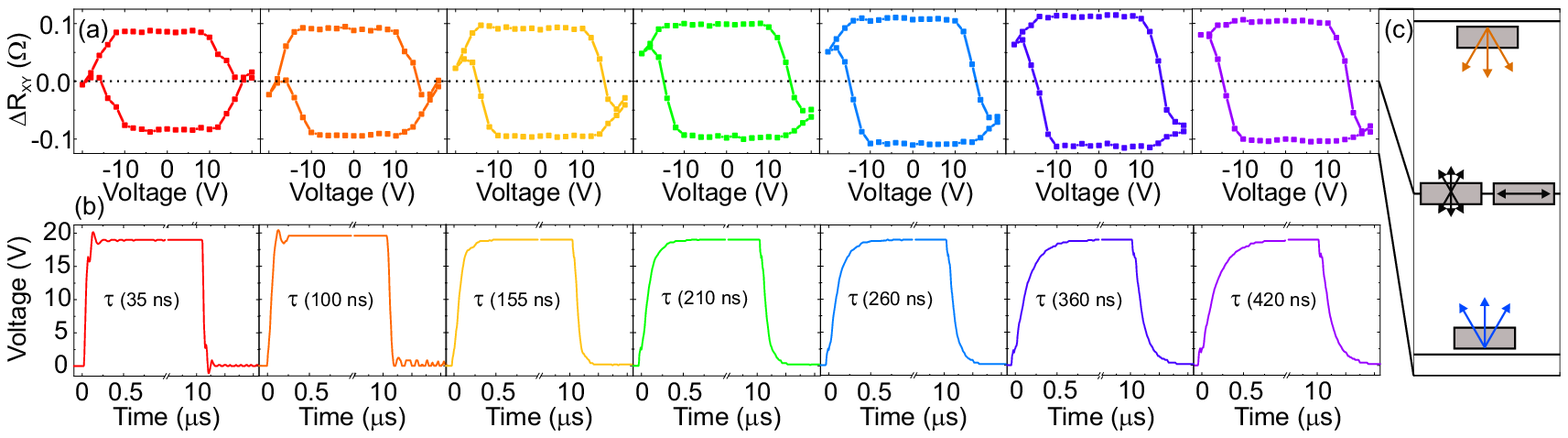}
	\label{fig:2}
	\caption{(a) Switching loops of Mn$_3$Sn/Pt as a function of pulse voltage with rise and fall time increasing from left to right. (b) Corresponding pulse shape. (c) Schematic showing the $-z$, $+z$ and intermediate states and the possible orientations of the \textbf{g}-vector in each state.}
\end{figure*}
\setlength{\textfloatsep}{10pt} 

Pioneering work on Mn$_3$Sn/heavy metal bilayers has demonstrated switching of antiferromagnetic order by current-induced spin-orbit torques \cite{Tsai2020, Tsai2021, Takeuchi2021, Deng2021}.
In these experiments, a change of the AHE as a function of current reveals the reorientation of \textbf{g} in crystal grains with $c$-axis oriented in-plane.
Switching only occurs in the presence of a symmetry-breaking magnetic field collinear with the current, with the final state determined by the relative orientation of current and field and by the sign of the spin Hall angle in the heavy metal \cite{Tsai2020, Tsai2021,Tsai2021a}. These observations suggest a switching mechanism very similar to ferromagnet/heavy metal bilayers \cite{Miron2011,Baumgartner2017,Manchon2019}. 
Within this picture, however, different magnetization dynamics can be expected depending on whether the torques rotate the moments in or out of the kagome plane \cite{Fujita2017,Yamane2019,Tsai2020,Takeuchi2021}. 
New effects such as chiral spin rotation have been proposed, whereby the Mn moments undergo continuous rotation in the kagome plane with time periods in the tens of ns \cite{Fujita2017,Takeuchi2021,Yan2022}. Thus far, however, switching experiments relied on electrical pulses with pulse duration of 100~ms, which yield no information about the fast switching dynamics expected of antiferromagnets.

In this work, we explore the chiral switching dynamics of Mn$_3$Sn/Pt bilayers. We observe that the switching behavior varies characteristically with the pulse length and shape: conventional bistable switching between $\pm z$ states is observed for pulses with fall times longer than 400~ns whereas tristable switching is observed for pulses with shorter fall times, leading to a demagnetized state with zero AHE. By studying the switching dependence on the temporal shape of the pulses, applied field, temperature, and time we show that the reversal of the \textbf{g}-vector occurs through two phases, namely current-induced partial demagnetization lasting several ns followed by cooling in the presence of spin-orbit torques at the end of a current pulse. This mechanism is similar to the setting of exchange bias during field cooling in coupled antiferromagnetic/ferromagnetic systems \cite{Nogues1999}. However, it differs from the thermally-activated switching observed in collinear ferromagnets \cite{Grimaldi2020} and antiferromagnets \cite{Meinert2018}, in which Joule heating reduces the magnetic anisotropy energy barrier while the sample remains magnetic. Time-resolved measurements during pulsing indicate that the reversal of chiral antiferromagnetic order is incoherent and that chiral spin rotation is either damped or averaged out in polycrystalline Mn$_3$Sn. Our measurements also set a limit on the reversal speed attainable by the interplay of current-induced heating and spin-orbit torques in chiral antiferromagnets. 

\section{Methods}
Our samples are polycrystalline Mn$_3$Sn(50~nm)/Pt(5~nm) bilayers grown by magnetron sputtering patterned into 3 to 6-$\mu$m-wide Hall bars and Hall crosses [Fig.~\ref{fig:1} (c,d)] \cite{Supmat}. High-resolution transmission electron microscopy reveals the presence of columnar grains of about 250~nm width, different orientations and excellent crystalline order \cite{Supmat}. Measurements of the longitudinal ($R_{\rm xx}$) and transverse Hall resistance ($R_{\rm xy}$) are consistent with previous work on similar samples \cite{Higo2018b,Tsai2020,Tsai2021,Tsai2021a,Yan2022,Supmat}. We used a quasi-static pulse-probe protocol for characterizing the switching properties as a function of pulse shape and field \cite{Miron2011} and performed the time-resolved measurements of the AHE using the split-pulse technique described in Ref.~\onlinecite{Sala2021}. In the pulse-probe method we inject a current pulse of up to 20~mA to induce switching followed by an alternate current of 1~mA, which allows for probing the first and second harmonic contributions to R\subt{xy} that are proportional to the AHE and ANE, respectively \cite{Supmat, Avci2014}. In the time-resolved measurements, we probe the change in AHE during a current pulse with about 50~ps temporal resolution \cite{Sala2021}. Hall bars are used for quasi-static switching and Hall crosses for the time-resolved measurements. Given the structure of our samples, the AHE (ANE) reflects the out-of-plane (in-plane) component of \textbf{g} averaged over different crystal grains in the region sensed by the Hall resistance \cite{Supmat,Higo2018a,Ikhlas2017,Yoon2020}. Comparative switching measurements on Mn$_3$Sn/W and W/Mn$_3$Sn/Pt samples are reported in Ref.~\onlinecite{Supmat}.

\section{Results}
\subsection{Multistate switching determined by the pulse fall time}
\begin{figure}[t]
	\includegraphics[width=1\columnwidth,height=\textheight,keepaspectratio]{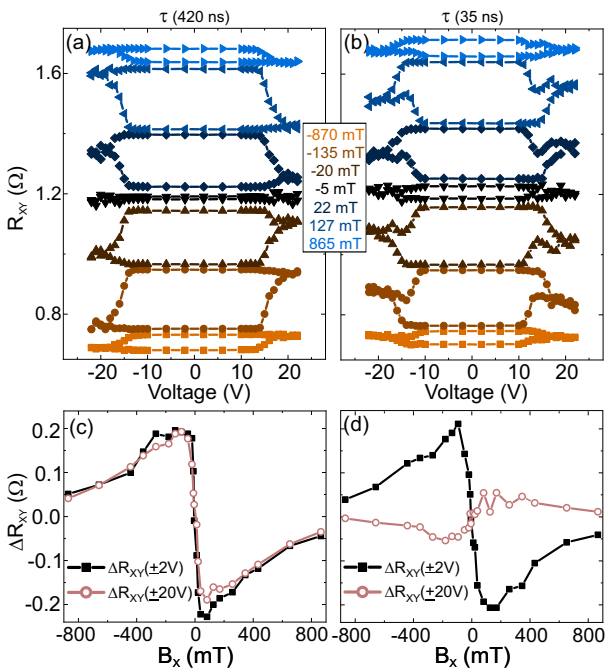}
	\label{fig:3}
	\caption{Field dependence of the current-induced switching for (a) long and (b) short fall time pulses of 10~$\mu$s length. Switching amplitude $\Delta R_{\rm xy}$ between $\pm2$V (black squares) and $\pm20$V (purple circles) as a function of $B_{\rm x}$ for (c) long and (d) short fall time.}
\end{figure}

Figures~\ref{fig:1} (e) and (f) show the field- and current-induced switching of the \textbf{g}-vector, respectively, as measured by the AHE. In agreement with previous reports \cite{Higo2018a,Tsai2020,Tsai2021}, we observe switching of about 30\% of the total AHE upon injecting 10-$\mu$s-long current pulses with a fall time $\tau = 420$~ns. This bistable behavior is interpreted as \textbf{g} switching between the $+z$ and $-z$ states. 
Surprisingly, however, we find that gradually reducing $\tau$ to below 100~ns changes the switching from bistable to tristable, leading to the appearance of states with high and low AHE at intermediate current values and zero AHE at high current [Fig.~\ref{fig:2}]. 
Because the pulse length is constant, the gradual shift of the endpoint $R_{\rm xy}$ in Fig.~\ref{fig:2} (a) demonstrates that the fall time determines the switching regime.  Importantly, the magnetic state set by the current pulse and magnetic field remains constant after the pulse.
The occurrence of multistate switching has been reported before in Mn$_3$Sn \cite{Takeuchi2021}, but the role of the transient dynamic effects that determine the final orientation of the \textbf{g}-vector has not been elucidated. These effects can be of two types, thermal, due to Joule heating, and magnetic due to spin-orbit torques.

\subsection{Switching as a function of in-plane field}

To exclude a purely thermal origin of the switching, we study its dependence on the external in-plane magnetic field $B_{\rm x}$. Figures~\ref{fig:3} (a) and (b) show the current-induced switching loops for 10-$\mu$s-long pulses with $\tau=35$~ns and $420$~ns, respectively, for increasing values of $B_{\rm x}$. The reversal of the switching direction upon inversion of $B_{\rm x}$ indicates that switching is due to spin-orbit torques in the entire range of fall times. We also find that the switching amplitude between $+z$ and $-z$ states, $\Delta R_{\rm xy}=R_{\rm xy}(2~{\rm V})-R_{\rm xy}(-2~\rm{V})$, increases up to $B_{\rm x}\approx 100$~mT, consistently with previous reports \cite{Tsai2020,Tsai2021,Tsai2021a,Higo2018a,Takeuchi2021} and the standard model of spin-orbit torque switching in ferromagnets \cite{Miron2011,Manchon2019}. However, $\Delta R_{\rm xy}$ decreases in the high field limit [black squares in Fig.~\ref{fig:3} (c,d)], indicating that another mechanism comes into play. We also note that the offset of $R_{\rm xy}$ and the sign of the switching amplitude $\Delta R_{\rm xy}(\pm \rm 20V)$ in the short pulse regime are very sensitive to the presence of an out-of-plane external field \cite{Supmat}.

\subsection{Switching by current-induced heating and cooling in the presence of spin-orbit torques}

\begin{figure}[t]
\includegraphics[width=1\columnwidth,height=\textheight,keepaspectratio]{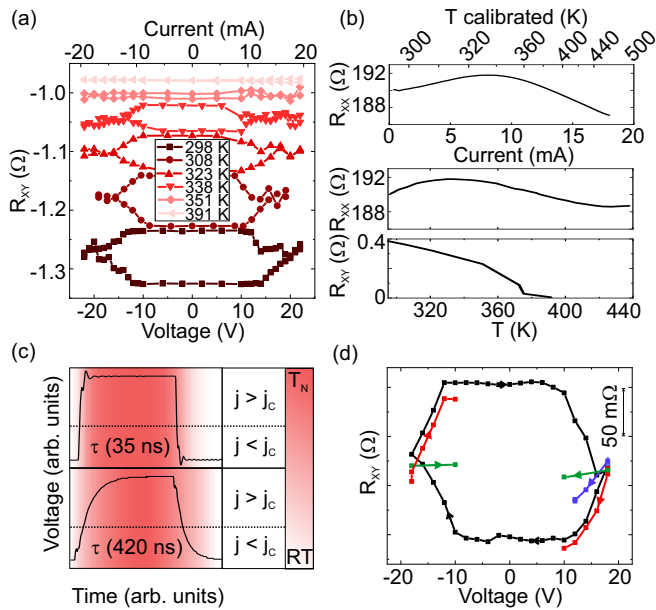}
	\label{fig:4}
	\caption{(a) AHE switching as a function of pulse voltage at different temperatures for 10-$\mu$s-long pulses. (b) Temperature dependence of $R_{\rm xy}$ (bottom panel) and $R_{\rm xx}$ (middle panel) of Mn$_3$Sn/Pt. $R_{\rm xx}$ vs direct current and calibrated temperature (top panel). (c) Schematic current pulse with temperature profile indicated by the red shading. 
	(d) Step-wise switching sequences: $R_{\rm xy}$ vs pulse amplitude starting from $\pm 18$~V in steps of 2~V (black, red), 4~V (blue) and 8~V (green).}
\end{figure}

To understand the role played by heating we measured the AHE as a function of temperature [Fig.~\ref{fig:4} (a,b)]. The AHE vanishes at $T_{\rm N}=390$~K, close to the Néel temperature of bulk Mn$_3$Sn (420~K) \cite{Song2020,Takeuchi2021}. The longitudinal resistance $R_{\rm xx}$ has a nonlinear temperature behavior as it is a mixture of the resistance due to Pt and Mn$_3$Sn. Measuring $R_{\rm xx}$ as a function of current allows us to gauge the extent of Joule heating, which shows that the sample temperature reaches $T_{\rm N}$ for pulse currents larger than 14~mA (16~V) \cite{Supmat}.
We thus propose a model to explain the multistate switching behaviour in which the interplay of temperature and spin-orbit torques is governed by $\tau$. Consider a generic voltage pulse that heats up the sample and provides a current density $j$ to exert a torque, as shown in Fig.~\ref{fig:4} (c). As the pulse starts, the temperature increases quadratically with the current at a rate determined by the longest between the pulse rise time and the heat diffusion time. For pulses longer than a few tens of ns, the sample temperature approaches $T_{\rm N}$, leading to a demagnetized state until cool down begins at the end of the pulse. Deterministic switching to a final state $+z$ or $-z$ can be achieved only if $j$ is larger than a critical current density $j_{\rm c}$ as the temperature has dropped below $T_{\rm N}$, i.e., for long $\tau$.  
If, on the other hand, the current drops abruptly below $j_{\rm c}$ when the temperature is still close to $T_{\rm N}$, the Mn$_3$Sn grains freeze in a mixed multidomain configuration, which leads to the intermediate state with no AHE for short $\tau$. Our simultaneous measurements of the AHE and ANE show that this intermediate state consists of domains along $\pm z$, which give a net zero AHE, and grains that are oriented along $+x$ and $-x$ for $B_{\rm x}>0$ and $B_{\rm x}<0$, respectively \cite{Supmat}. 
The fraction of grains oriented along $\pm x$ during cool down increases with $B_{\rm x}$, which explains the non-monotonic field dependence of the switching amplitude in Fig.~\ref{fig:3}. Thus, by gradually modifying $\tau$, we tune the fraction of grains that switch and those that remain demagnetized at the end of the pulse.

This model also explains why the $\pm z$ final states with large/low AHE can be reached starting from the intermediate state with zero AHE upon reducing the pulse voltage in small incremental steps, as seen in Fig.~\ref{fig:2} even for short $\tau$. 
Figure~\ref{fig:4} (d) shows $R_{\rm xy}$ recorded by sweeping the pulse amplitude from +18~V to -18~V and back in steps of 2~V (black dots). Starting from the intermediate state obtained by pulsing at +18~V with $\tau=35$~ns, the AHE changes progressively to the low state upon reducing the pulse amplitude. 
However, if the pulse amplitude is abruptly decreased from +18 to +10~V, no switching occurs (green dots). The type of switching thus depends on the initial state and on the decremental step size, which is different from the change of switching amplitude as a function of current reported for bistable switching in Ref.~\onlinecite{Tsai2020}. Our observation is consistent with different Mn$_3$Sn grains having a distribution of $T_{\rm N}$ due to their varying sizes, which are selectively switched to the $\pm z$ final states upon decreasing the pulse amplitude from the intermediate state. This is essentially a step-wise version of the long fall time scenario described above.

Overall, our results show that the switching of antiferromagnetic order in Mn$_3$Sn occurs due to heat-assisted demagnetization followed by reorientation of the \textbf{g}-vector induced by spin-orbit torques during cool down. The fall time of the current pulses determines the final magnetic configuration of the Mn$_3$Sn domains. Additionally, switching loops measured for 21~V pulses of decreasing length, from 50 to 5~ns, evidence that the switching amplitude vanishes in the limit of short pulses [Fig.~\ref{fig:5} (a)]. These findings show that switching of antiferromagnetic order in Mn$_3$Sn by spin-orbit torques has a composite temporal dependence and a different dynamics relative to ferromagnets \cite{Sala2021,Krizakova2021} and collinear antiferromagnets \cite{Meinert2018,Wornle2019,Kaspar2021}.

\subsection{Time-resolved measurements}

\begin{figure}[!ht]
	\includegraphics[width=1\columnwidth,height=\textheight,keepaspectratio]{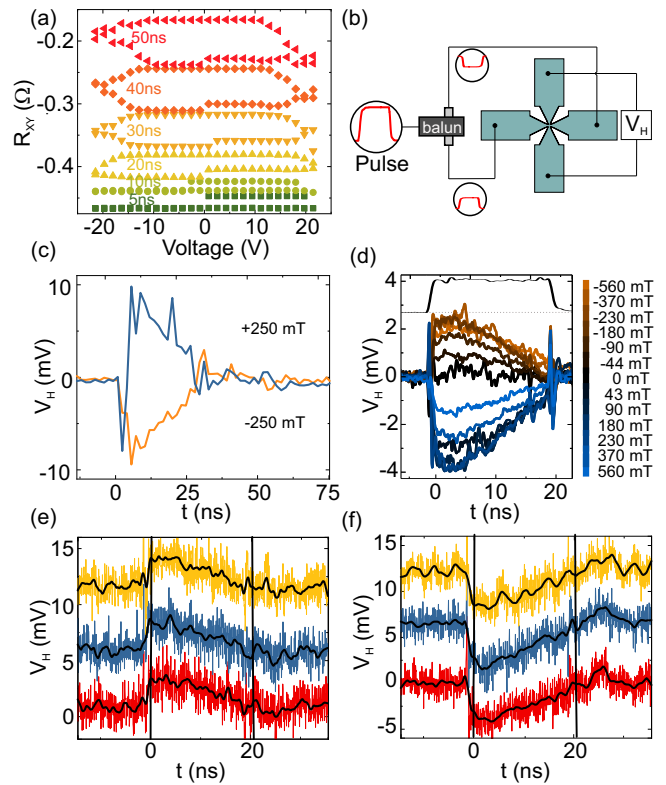}
	\label{fig:5}
	\caption{(a) Current-induced switching loops for different pulse lengths, $\tau=0.3$~ns and $B_{\rm x}=- 250$~mT. (b) Schematic of the time-resolved AHE measurements.
	(c) Differential switching time traces averaged over 20 consecutive +21~V pulses with $B_{\rm x}=\pm 250$~mT. The pulses are 75~ns long starting at $t=0$. (d) Differential switching time traces averaged over 100 consecutive 20-ns-long voltage pulses of amplitude +21~V vs $B_{\rm x}$. The gray trace at the top shows the pulse shape. 
	(e) Single shot differential switching traces for 20-ns-long voltage pulses at $B_{\rm x}=-180$~mT and (f) +180~mT. The black lines are moving averages over $1.5$~ns.}
\end{figure}

To determine the transient dynamics, we performed time-resolved measurements of $R_{\rm xy}$ during the current pulses using the setup shown in Fig.~\ref{fig:5} (b). The temporal evolution of the AHE voltage $V_{\rm H}$ during the switching process is determined by taking the difference of the Hall voltage trace measured during switching relative to a reference trace in the absence of switching \cite{Sala2021}. Figure \ref{fig:5} (c) shows the average of 20 differential time traces of $V_{\rm H}$ taken during pulses with +21~V amplitude, 75~ns duration and $\tau=0.3$~ns, separated by a 1~s delay. The decrease (increase) of $V_{\rm H}$ following the onset of the pulse at $t=0$ for $B_{\rm x}=+ 250$~mT ($- 250$~mT) reflects the decrease (increase) of the AHE from the initial $-z$ ($+z$) state to the intermediate state with no AHE. 
It takes about 35~ns for $\lvert V_{\rm H} \rvert$ to reduce to 0, after which no further changes of $V_{\rm H}$ are observed until the end of the pulse. Measurements performed for 20~ns-long pulses as a function of $B_{\rm x}$, reported in Fig.~\ref{fig:5} (d), further reveal that the amplitude of the transient switching signal scales with $B_{\rm x}$ and that the timescale over which $\lvert V_{\rm H} \rvert$ reduces to 0 is independent of $B_{\rm x}$. We thus associate the decrease of $\lvert V_{\rm H} \rvert$ with the time it takes for the device to reach a temperature close to $T_{\rm N}$, in line with the switching mechanism proposed above. This time depends only on $j$ and not on $B_{\rm x}$, which shows that the switching speed of Mn$_3$Sn is ultimately limited by the heating rate.

Recent studies propose a coherent chiral spin reversal mechanism in noncollinear antiferromagnets where the \textbf{g}-vector continuously rotates above a given current density threshold \cite{Fujita2017,Takeuchi2021,Yan2022}. The rotation period is estimated in a range of 1-30~ns, depending on the current density. Inirect evidence for this effect has been reported both in epitaxial and polycrystalline thin films \cite{Takeuchi2021,Yan2022}. 

The experimental evidence for such a mechanism, however, lacks insight into the time-dependent dynamics that is the hallmark of coherent switching. Our time-resolved traces shown in Fig.~\ref{fig:5} (c,d) evidence a monotonic decrease of $\lvert V_{\rm H} \rvert$ that is not consistent with reproducible oscillations of $R_{\rm xy}$ due to chiral spin rotation. Because these traces are averaged over several pulses, they do not provide information on stochastic rotations. To investigate the occurrence of chiral spin rotation during individual current pulses, we have thus measured single-shot time traces of $V_{\rm H}$. Representative examples of such traces are shown in Fig.~\ref{fig:5} (e,f) for a series of 20-ns-long current pulses. Our analysis does not reveal evidence of periodic oscillations of $V_{\rm H}$ consistent with chiral spin rotation during single-shot pulses. 

The absence of oscillations can be ascribed to different factors. First, chiral spin rotation requires an injected spin current with polarization parallel to the $c$-axis \cite{Takeuchi2021}. Given the polycrystalline nature of our samples, we estimate to have a measurable amount of such grains in our Hall crosses \cite{Supmat, Takeuchi2021, Yan2022}. On the other hand, chiral spin rotation may take place in different grains with different phase factors, averaging to zero in the total Hall signal. Acording to simulations, however, these coherent effects should result in visible oscillations also in polycrystalline samples \cite{Yan2022}. Another possibility is that the rotation is too fast to be resolved by our measurements, which have a temporal resolution of about 50~ps \cite{Sala2021}. The current density in the time-resolved measurements is $1.3\times10^7$~A/cm$^2$ when averaged over the entire thickness of Mn$_3$Sn/Pt and about $6.2\times10^7$~A/cm$^2$ in Pt, as estimated using a parallel resistor model. The rotation frequency corresponding to this current is 1.7~GHz \cite{Takeuchi2021}, which is within our time resolution. 
The continuous decrease of the AHE signal thus indicates that any oscillation, if present, is strongly damped and that heat-induced demagnetization dominates over coherent effects. 

\section{Conclusions}
In summary, our work shows that the switching of chiral antiferromagnetic order in Mn$_3$Sn/Pt is incoherent and determined by the timed interplay of heat and spin-orbit torques. Both effects are current-induced but heating up to $T_{\rm N}$ occurs on a timescale of tens of ns whereas the injection of a spin current from Pt closely follows the temporal profile of the current pulses. Switching proceeds via a two-step demagnetization-remagnetization process, whereby the final orientation of the \textbf{g}-vector is deterministic between $\pm z$ states only if the sample cools down in the presence of a spin current larger than a critical value. Our results provide insight into the switching timescale and dynamics of topological antiferromagnets, showing that it is different from both ferromagnets and collinear antiferromagnets and limited by the sample heating rate. 
Additionally, this work shows that time-resolved Hall effect measurements provide a viable method to investigate the current-induced dynamics of antiferromagnetic order in topological materials.

\emph{Acknowledgements}. This research was partially supported by the Swiss National Science Foundation (Grants No.
200020-200465 and PZ00P2-179944). During the resubmission of this manuscript we became aware of related work reporting the multistate switching of Mn$_3$Sn \cite{Pal2022}.

\section{References}
\bibliography{Mn3Sn_paper_2}
\end{document}


\title{Supplemental Material\\ Time-dependent multistate switching of topological antiferromagnetic order in Mn$_3$Sn}
	\author{Gunasheel Kauwtilyaa Krishnaswamy}
	\affiliation{Department of Materials, ETH Zurich, 8093 Zurich, Switzerland}
	\author{Giacomo Sala}
	\affiliation{Department of Materials, ETH Zurich, 8093 Zurich, Switzerland}
	\author{Benjamin Jacot}
	\affiliation{Department of Materials, ETH Zurich, 8093 Zurich, Switzerland}
	\author{Charles-Henri Lambert}
	\affiliation{Department of Materials, ETH Zurich, 8093 Zurich, Switzerland}
	\author{Richard Schlitz}
	\affiliation{Department of Materials, ETH Zurich, 8093 Zurich, Switzerland}
	\author{Marta D. Rossell}
	\affiliation{Electron Microscopy Center, Empa, Swiss Federal Laboratories for Material Science and Technology, Dübendorf, Switzerland}
	\author{Paul Nöel}
	\affiliation{Department of Materials, ETH Zurich, 8093 Zurich, Switzerland}
	\author{Pietro Gambardella}
	\affiliation{Department of Materials, ETH Zurich, 8093 Zurich, Switzerland}
	
	\maketitle 
	\tableofcontents
	\date{\today}
	\cleardoublepage
	
	\section{SM1. Sample fabrication}
	
	Mn$_3$Sn is widely known to exhibit a large anomalous Hall effect (AHE) for an antiferromagnet. However, in most cases of current-induced switching of Mn$_3$Sn, the switching amplitude does not exceed 50\% of the total AHE resistance.
	Switching amplitudes of few tens of milliohms are usually shown for 100-ms-long current pulses. On further reducing the pulse widths to few nanoseconds, the typical switching amplitude diminishes to much smaller Hall signals. Time-resolved measurement of the AHE in the nanosecond regime in practice requires at least 40~m$\Omega$ of switching amplitude. 
	The latter depends on three main ingredients, namely: maximum AHE resistance \cite{Tsai2021a}, current shunting due to buffer layers and heavy metal layer, interfacial quality for spin transmission from the heavy metal. The AHE resistance is tuned by the stoichiometry \cite{Ikeda2019, Ikeda2020, Yoon2020,Deng2021, Yoon2021} and the largest amplitude is obtained when $\delta =0$ for Mn$_{3+\delta}$Sn. The shunting resistance can be reduced by reducing the thickness of buffer layers such as Ru and heavy metals such as Pt and W. The interfacial quality that leads to the best switching is obtained for a slight intermixing of the atoms of the interface \cite{Tsai2021, Tsai2021a}. Pt tends to form alloys of Mn$_{\rm x}$PtSn whereas W is more resilient to intermixing. Taking all these considerations into account we prepared different Mn$_3$Sn samples with Pt and W as heavy metals without buffer layers for current-induced switching, as well as control samples of Mn$_3$Sn grown with and without a Ru buffer layer.
	
	The samples were grown by DC magnetron sputtering on SiO$_2$ substrates. The substrates were sonicated in acetone and isopropanol for 10 minutes each before deposition. Mn\subt{3}Sn was deposited at room temperature (RT) by co-sputtering of a MnSn and Mn target with powers of 13~W and 9~W, respectively, and an Ar pressure of 3.2~mTorr at a growth rate of 0.05~nm/s. 
	The sample stoichiometry was calibrated by scanning electron microscopy-energy dispersive x-ray spectroscopy and independently with Rutherford backscattering spectrometry. The sample thickness was calibrated by atomic force microscopy on amorphous Mn$_3$Sn devices fabricated by lift-off. Pt, W and Ru were deposited with a nominal power of 6~W at 3.2~mTorr Ar pressure at rates of 0.076~nm/s, 0.026~nm/s and 0.029~nm/s, respectively. The Si$_3$N$_4$ capping was deposited by RF sputtering at 50~W and 8.5~mTorr of Ar at a rate of 0.029~nm/s. Hall bars with $16~\mu$m length and $\times 3.7\mu$m width and Hall crosses with width varying from $2.2~\mu$m to $6.6~\mu$m were patterned by UV photolithography and reactive ion etching.
	
	The sample with the largest AHE reported in the main text is SiO$_2$/Mn$_3$Sn(50nm)/Pt(5nm) /Si$_3$N$_4$(5nm), here referred to as sample \#85 (s85). The Mn$_3$Sn layer was grown at RT and annealed at 600°C for 1 hour in vacuum, and Pt and Si$_3$N$_4$ were deposited after annealing during the cooldown to RT at a rate of 5~K/min. This was done to induce slight intermixing at the interface to ensure a large switching amplitude of the AHE as discussed in the literature \cite{Tsai2021a}. 
	
	Switching experiments were also performed on SiO$_2$/Mn$_{3.16}$Sn(50nm)/W(4nm)/ Ru(1nm) (s41) and SiO$_2$/W(4nm)/Mn$_3$Sn(50nm)/Pt(4nm) (s90). s41 was grown at RT and annealed at 550°C for 1h, then cooled down at a rate of 5~K/min. W and Ru were grown at RT after annealing. This sample has a nominal excess of Mn to avoid Mn deficient spurious phases. In s90 the W and Mn$_3$Sn layers were deposited at 420°C and annealed at 600°C for 1h and cooled down at a rate of 5~K/min. Pt was deposited at RT after cooling. 
	
	\begin{figure}
		\includegraphics[width=0.5\columnwidth,height=\textheight,keepaspectratio]{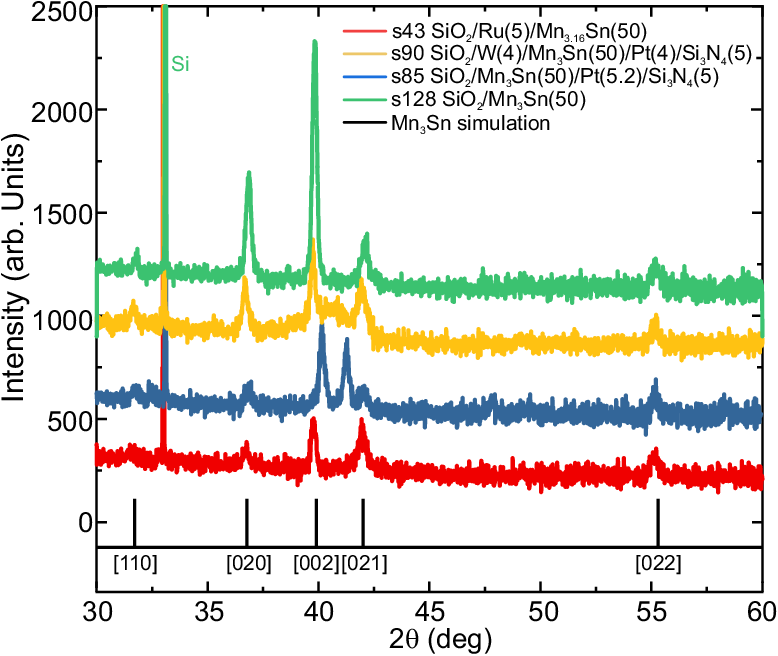}
		\label{fig:6}
		\caption{X-ray diffraction of Mn$_3$Sn polycrystalline samples grown under different conditions.}
		\end{figure}
		Figure~\ref{fig:6} compares the x-ray diffraction (XRD) spectra of s85, s90, and two control samples, namely SiO$_2$/Ru(5nm)/Mn$_{3.16}$Sn(50nm) (s43) 
		and SiO$_2$/Mn$_{3.16}$Sn(50nm) (s128), both grown in the same conditions as s85 and without capping layer. 
		All the samples are polycrystalline with grains of different orientation. s85 presents an additional peak between the [002] and [021] reflections that we attribute to the formation of crystalline Si$_3$N$_4$ after annealing. XRD measurements performed on patterned devices from sample s85 confirm the absence of spurious phases, in agreement with the transmission electron microscopy data reported in Sect. SM2. s90 presents a broad structure around $2\theta=$41° which is due the growth of highly textured W at 400~°C. The samples s43 and s128 present the same [020], [001], [021], [022], [110] peaks as s85 and s90, albeit with different intensity as observed in literature depending on the growth conditions \cite{Higo2018b,Ikeda2019,Yoon2020,Yoon2021,Tsai2021a,Nakano2021}.
		
		The anomalous Hall resistivity of s85 is $2~\mu\Omega\rm cm$, in agreement with that of polycrystalline films prepared for switching experiments \cite{Tsai2020,Tsai2021,Tsai2021a}, and the current-induced switching amplitude is about 250-300~m$\Omega$ for 10-$\mu$s-long current pulses, depending on the device. The anomalous Hall resistivity of s43 is $1 \mu\Omega\rm cm$, and the switching amplitude $45~\rm m\Omega$ comparable in magnitude to Ref.\cite{Tsai2020}. The anomalous Hall resistivity of s90 is $0.8~\mu\Omega\rm cm$, and the switching amplitude $74~\rm m\Omega$. All the samples present a negative AHE resistance and large out-of-plane coercivity, as expected of Mn$_{3}$Sn.

	\section{SM2. Transmission electron microscopy of Mn$_3$Sn/Pt}
		
		\begin{figure}[h]
			\centering
			\includegraphics[width=\textwidth]{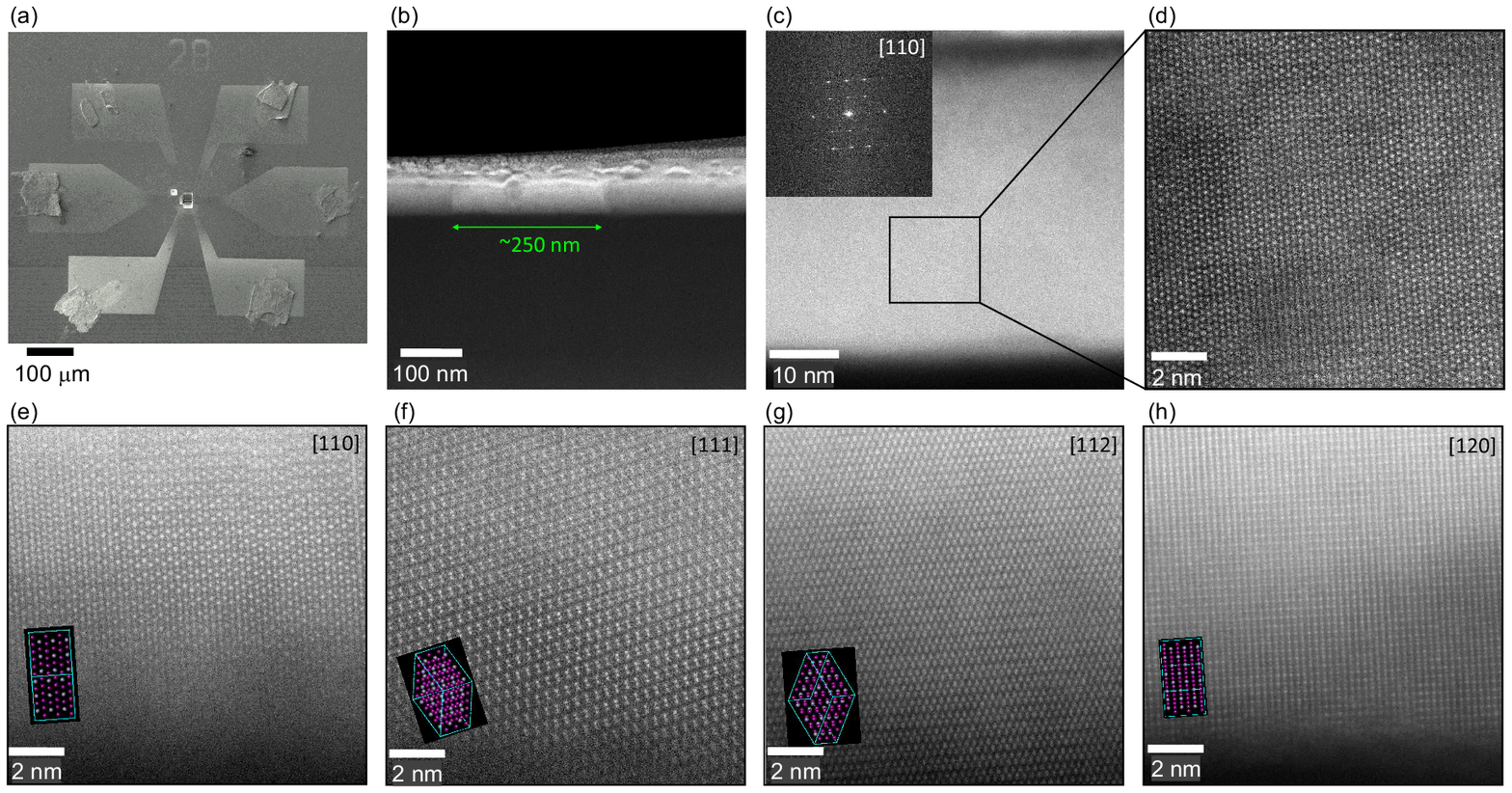}
			\caption{(a) Scanning electron micrograph of a device from which a lamella for STEM imaging was excavated. The device belongs to sample s85. (b) HAADF-STEM image showing columnar grain growth with grains extending through the thickness of the film. The brighter 250 nm wide grain is oriented along the [110] zone axis. (c) Detail of the [110]-oriented grain showing well-ordered lattice planes as seen in the magnified image of panel (d). The inset shows the corresponding FFT pattern that can be assigned to the Mn$_3$Sn phase. (e-h) HAADF-STEM images of grains with different crystallographic orientations observed in the same film. The insets are simulated crystal structures. Purple and silver spheres correspond to Mn and Sn atoms, respectively.}
			\label{fig:14}
		\end{figure}
		
		We performed transmission electron microscopy (TEM) measurements to inspect the crystallographic grain structure present in the polycrystalline film. An electron-transparent lamella for TEM investigation was excavated from sample s85 using a FEI Helios 660 G3 UC focused ion beam after deposition of a Pt protective layer. The lamella was extracted from the centre of the Hall bar, as can be seen from the milled area shown in Fig. \ref{fig:14} (a). A FEI Titan Themis operated at 300 kV was used for high-angle annular dark-field scanning transmission electron microscopy (HAADF-STEM) imaging (probe semi-convergence angle: 18 mrad, annular semi-detection range: 66-200 mrad). A HAADF-STEM image of a region of the film revealing two distinct grain boundaries is shown in Fig.\ref{fig:14} (b). The film exhibits a highly columnar microstructure, with grain widths greater than 250 nm extending from the substrate to the top surface. Details of the brighter [110]-oriented grain are shown at higher magnifications in Figs. \ref{fig:14} (c,d). The inset in panel (c) shows the corresponding fast fourier transform (FFT) pattern with sharp spots indicative of a single crystal. Figures \ref{fig:14} (e-h) show atomically resolved HAADF-STEM images of other film regions with crystal grains oriented along the [110],[111], [112] and [120] zone axes, respectively. The insets are magnified images with overlaid simulated crystal structures. The HAADF-STEM images confirm the high quality of our polycrystalline Mn$_3$Sn devices.
		
	\section{SM3. Quasi-static and time-resolved measurements of the anomalous Hall resistance}
		
		We performed two types of measurements of R$_{\rm xy}$: quasi-static switching hysteresis loops [Fig 1~(f), 2~(a), 3~(a,b), 4~(a,d), 5~(a)] and time-resolved measurements [Fig 5 (c-f)] in the main text. The quasi-static measurements are performed using a pulse-probe scheme as mentioned in the manuscript. A voltage pulse of amplitude $V$, length $t_{\rm p}$ ranging from  35~ns to 10~ms, and rise/fall time ranging from $\tau = 35$~ns to $420$~ns is applied to the sample for switching. 0.1~s after the end of the pulse, we apply an alternate (ac) current of amplitude 1~mA at a frequency of 10 Hz to probe the anomalous Hall resistance $R_{\rm xy}$. The $R_{\rm xy}$ is acquired for 0.5~s by an analog-to-digital converter that separates the first and second harmonic resistance signals. The next pulse is sent immediately after the acquisition ends. By changing the value of $V$ after each pulse, we construct the switching hysteresis loop as a function of voltage (or current). The anomalous Hall resistance is given by the first harmonic $R_{\rm xy}$. To measure the anomalous Nernst resistance we take the second harmonic $R_{\rm xy}$ and use ac currents of 8~mA during the probing. This current is high enough to induce a large magnetothermal effect but low enough not to affect the magnetic state of the sample.
		
		Time-resolved measurements, instead, probe $R_{\rm xy}$ during the pulse using the current of the pulse itself to measure the Hall voltage. Here we use a set-reset scheme to measure the switching: Starting from the initial state, which can be determined by field or current, we send a pulse of length $t_{\rm p}= 5-100$~ns, rise/fall time $\tau = 0.3$~ns, and amplitude +20~V during which we measure the time-dependent Hall voltage at the sampling rate of our oscilloscope (20~GS/s). After 5~s, we apply a reset pulse of -20~V to re-initialise the system in the opposite polarity state. Single-shot measurements report the time-dependent changes of the Hall voltage during one pulse. We also average consecutive single-shot measurements to achieve a better signal-to-noise ratio. 
		
		We use both Hall bars and Hall crosses for quasi-static measurements but only Hall crosses for the time-resolved measurements. In the latter, one pulse is split into two halves which are counter propagating along the current line to impose a virtual ground at the middle of the Hall cross where the Hall voltage is measured. In order to synchronise the counter-propagating pulses, we use Hall crosses for which the RF cabling lengths and circuitry are optimised to minimize spurious signals. The details of how this technique works are presented in a previous publication \cite{Sala2021}.
		
	\section{SM4. Current-induced switching of Mn$_3$Sn/W and W/Mn$_3$Sn/Pt}
		\begin{figure}[h]
			\includegraphics[width=\columnwidth,height=\textheight,keepaspectratio]{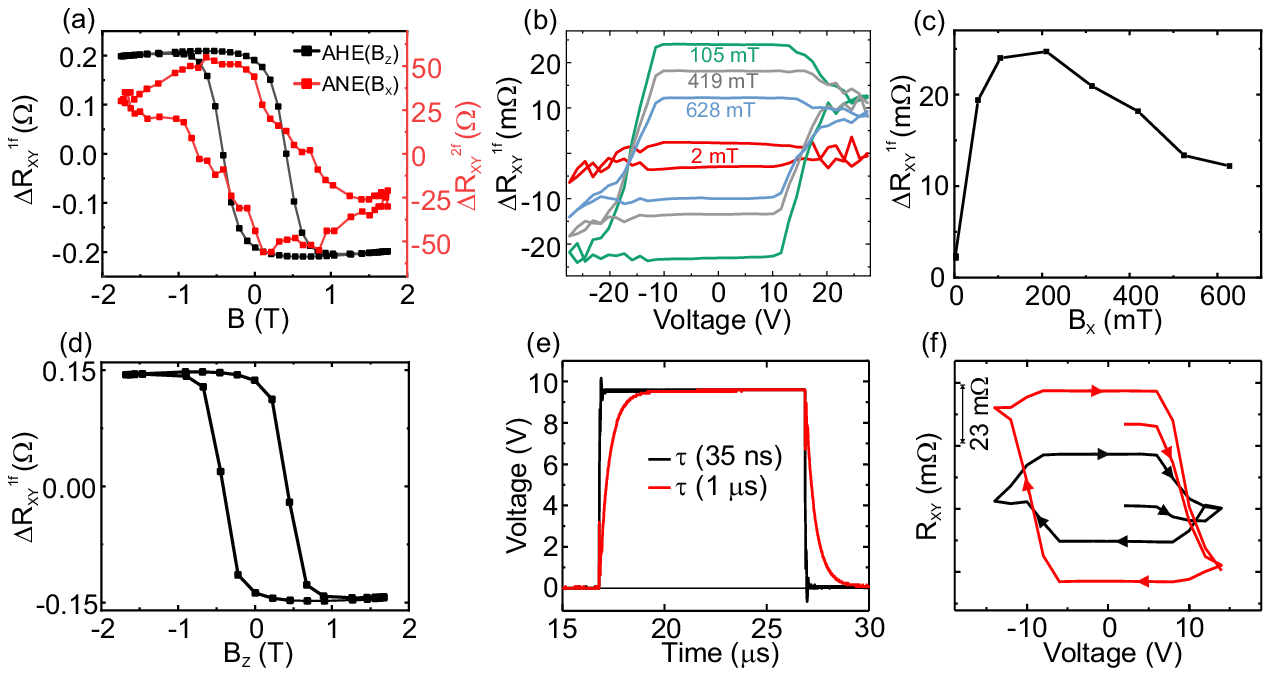}
			\label{fig:7}
			\caption{(a) AHE and ANE of SiO$_2$/Mn$_{3.16}$Sn(50nm)/W(4nm)/Ru(1nm) (s41) as a function of out-of-plane field $B_{\rm z}$ and in-plane field $B_{\rm x}$, respectively. (b) Switching of the AHE with 10-$\mu$s-long current pulses with rise/fall time $tau=2~\mu$s  for different values of $B_{\rm x}$. (c) Dependence of the switching amplitude on $B_{\rm x}$. (d) AHE of SiO$_2$/W(4nm)/Mn$_3$Sn(50nm)/Pt(4nm) (s90) as a function of $B_{\rm z}$.
				(e) Pulse shape for $\tau=35$~ns and $1~\mu$s. (f) Characteristic switching loops pulses with $\tau=35$~ns (black line) and $1~\mu$s (red line).}
		\end{figure}
		Figure~\ref{fig:7} (a) shows the AHE and anomalous Nernst effect (ANE) of a sample where W replaces Pt as the spin current source, namely SiO$_2$/Mn$_{3.16}$Sn(50nm)/W(4nm)/Ru(1nm) (s41). The AHE and ANE are both negative and have a large coercivity similar to sample s85 described in the main text. The current-induced switching and its dependence on the in-plane magnetic field $B_{\rm x}$ show the characteristic behavior with a peak in switching amplitude around 100~mT and the subsequent suppression on further increase of $B_{\rm x}$, as observed in Fig.~\ref{fig:7} (b) and (c). The switching polarity is positive for W deposited on top due to the negative spin Hall angle of W. 
		
		Figure~\ref{fig:7} (d) shows the AHE of SiO$_2$/W(4nm)/Mn$_3$Sn(50nm)/Pt(4nm) (s90), which is similar in magnitude and coercivity to that of s41.  Figure~\ref{fig:7} (e,f) show the shape of the voltage pulses with $\tau=35$~ns and 420~ns used for switching and the corresponding switching loops as a function of pulse amplitude. The switching behavior for pulses with long and short fall time is analogous to that reported for SiO$_2$/Mn$_3$Sn(50nm)/Pt(5nm)/Si$_3$N$_4$(5nm) (s85) in the main text.

	\section{SM5. Temperature calibration}
\begin{figure}[h]
			\includegraphics[width=\columnwidth,height=\textheight,keepaspectratio]{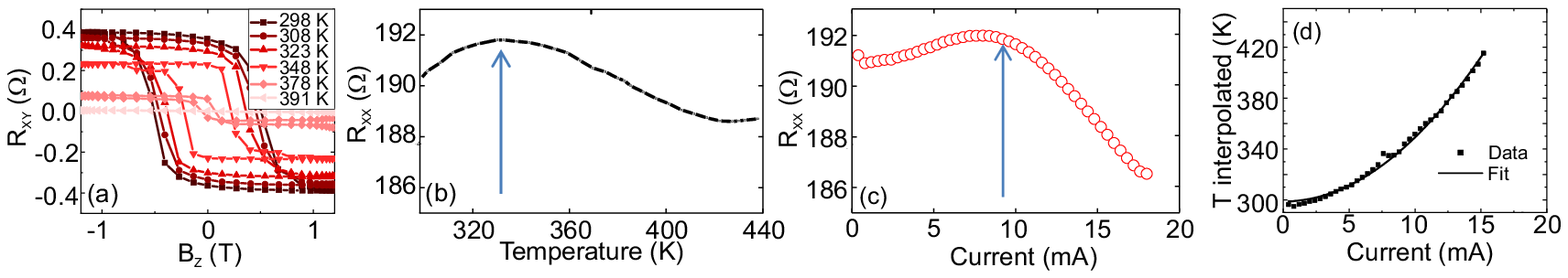}
			\label{fig:8}
			\caption{(a) $R_{\rm xy}$ and (b) $R_{\rm xx}$ of SiO$_2$/Mn$_3$Sn(50nm)/Pt(5nm)/Si$_3$N$_4$(5nm) as a function of temperature. (c) $R_{\rm xx}$ as a function of current in ambient conditions.
				The blue arrow in (b) and (c) is indicative of the change in resistance regime from Pt-dominated to Mn$_3$Sn-dominated.(d) Temperature obtained from the piece-wise linear interpolation of $R_{\rm xx}$ vs current. The line is a quadratic fit to the data.}
		\end{figure}

		Figure~\ref{fig:8} (a) shows the AHE resistance $R_{\rm xy}$ of SiO$_2$/Mn$_3$Sn(50nm)/Pt(5nm)/Si$_3$N$_4$(5nm) (s85) as a function of out-of-plane field measured at different temperatures, which is consistent with a Néel temperature $T_{\rm N} \approx 390$~K. To calibrate the sample temperature as a function of current, we measured the longitudinal resistance $R_{\rm xx}$ during a current sweep and compared it to the temperature dependence of $R_{\rm xx}$, as shown in Fig.~\ref{fig:8} (b,c). Interpolating the $R_{\rm xx}$ vs current curve gives the temperature vs current characteristic shown in Fig.~\ref{fig:8} (d). A clear quadratic relationship between temperature $T$ and current $I$ is obtained, given by $T= 298.7 + 0.50 \cdot I^2$, as expected for Joule heating. This calibration shows that the sample temperature overcomes $T_{\rm N}$ for typical switching currents between 15~mA and 20~mA.
		
		We compare below our temperature calibration with that obtained by other studies.  
		
		The samples of Ref.~\onlinecite{Tsai2020} are comparable to ours in terms of thickness. A temperature rise of 50 K from room temperature is mentioned, but it is not specified how this was measured or estimated. These samples switch at a critical current of 40 mA, corresponding to a current density of $7 \times 10^6$~A/cm$^2$ and power of 0.32 W. The critical current for switching, however, increases by 10 mA when the measurements are performed at 200 K compared to 300 K. Our measurements show that even 10 mA difference can lead to a substantial change in temperature. 
		
		Reference~\onlinecite{Takeuchi2021} reports post-pulse AHE measurements of 8.3-nm-thick Mn$_3$Sn and a temperature calibration performed on 9.9-nm-thick Mn$_3$S. The latter samples reach a temperature above 400~K close to $T_{\rm N}$ at a current of 18 mA as used for bipolar switching.  An increase in AHE noise is reported starting from 15 mA in the bipolar regime. The AHE fluctuations vs current and pulse number are interpreted as evidence for chiral spin rotation. For a 10-nm-thick Mn$_3$S film with 5-nm-thick Pt and a device size of $9.8 \times 43$~$\mu$m$^2$, the calculated current density for 18 mA is $1.03 \times 10^7$~A/cm$^2$ and the power is 0.12~W. 
		
		Reference~\onlinecite{Pal2022} reports post-pulse AHE measurements of 30 to 100~nm-thick Mn$_3$Sn with 8-nm-thick W. They estimate a temperature rise of 140~K during a 1-ms-long pulse at saturation current densities of $1.7 \times 10^7$~A/cm$^2$ by measuring the resistance offset of their devices as a function of temperature. At the critical switching current density of $1.2 \times 10^7$~A/cm$^2$, the dissipated power is 0.2~W.

		Our critical switching current density is $7.5 \times 10^6$~A/cm$^2$ for 10-$\mu$s-long pulses in s85, corresponding to a power of 0.26 W. These values are comparable to those employed in other studies of Mn$_3$Sn switching. 
		
		We also note that the power dissipated by Joule heating scales proportionally with the sample volume and the heat conduction to the substrate and air is worse in thicker films. Thus typically more heat is generated in thicker films compared to thinner films for the same current density, which complicates the comparison between different samples.

	\section{SM6. Orientation of domains in the intermediate state probed by the anomalous Nernst effect}
		
	To investigate the orientation of the \textbf{g}-vector in the intermediate state, we performed simultaneous measurements of the AHE and ANE using a large ac sensing current of 8~mA and detecting the first and second harmonic components of the Hall resistance, $R_{\rm xy}^{\rm 1f}$ and $R_{\rm xy}^{\rm 2f}$, respectively. Here $R_{\rm xy}^{\rm 1f}$ measures the AHE, which is linear with the current, and $R_{\rm xy}^{\rm 2f}$ measures the ANE, which is a thermal effect that scales quadratically with the current \cite{Avci2014}.
		\begin{figure*}
			\includegraphics[width=0.5\textwidth,keepaspectratio]{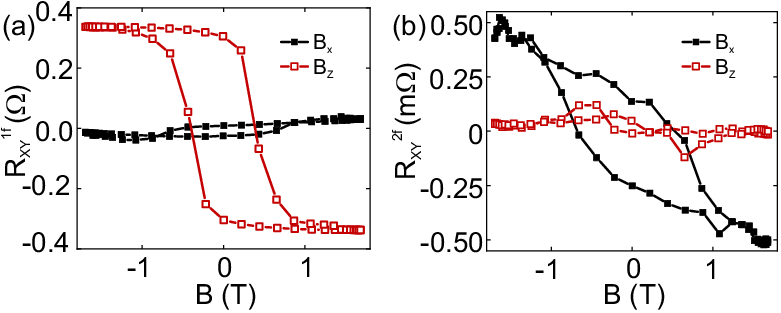}
			\label{fig:10}
			\caption{(a) AHE and (b) ANE of SiO$_2$/Mn$_3$Sn(50nm)/Pt(5nm)/Si$_3$N$_4$(5nm) (s85) probed as a function of $B_{\rm z}$ (red open squares) and $B_{\rm x}$ (black filled squares) to simultaneously measure the $z$ and $x$ components of the \textbf{g}-vector.}
		\end{figure*}
		
		Figure~\ref{fig:10} (a,b) shows the AHE and ANE of SiO$_2$/Mn$_3$Sn(50nm)/Pt(5nm)/Si$_3$N$_4$(5nm) (s85) measured as a function of $B_{\rm z}$ (red open squares) and $B_{\rm x}$ (black filled squares). The AHE provides information on the $z$ component of the \textbf{g}-vector, showing that $g_{\rm z}$ aligns with $B_{\rm z}$, as expected, whereas $g_{\rm z}\approx 0$ as a function of $B_{\rm x}$. The ANE, on the other hand, is proportional to $\textbf{g}\times\nabla T_{\rm z}$, where $\nabla T_{\rm z}$ is the out-of-plane thermal gradient generated by Joule heating. Thus the ANE generates a thermoelectric voltage along $y$ when \textbf{g} is oriented along $x$. The measurements in Fig.~\ref{fig:10} (b) show that \textbf{g} has a large in-plane component $g_{\rm x}$ that aligns with $B_{\rm x}$, as expected for a polycrystalline Mn$_3$Sn layer with different grain orientations.
		
		\begin{figure*}[t!]
			\includegraphics[width=\textwidth,keepaspectratio]{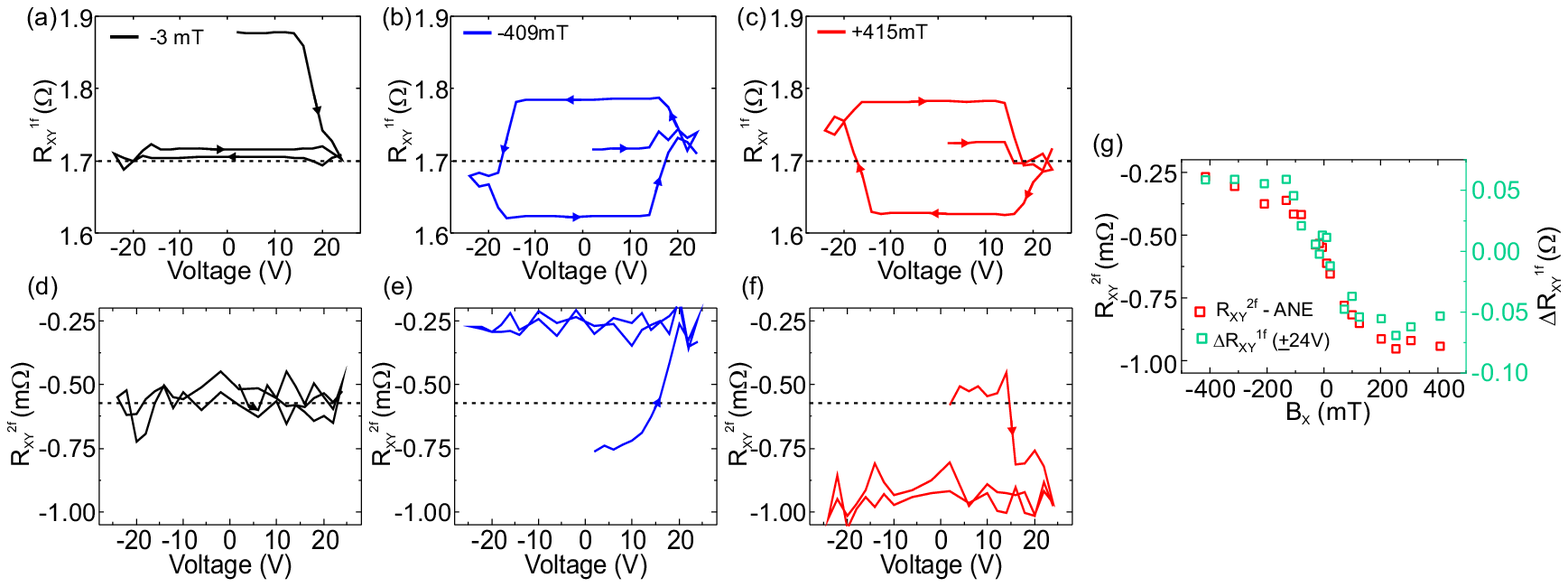}
			\label{fig:11}
			\caption{Switching loops as a function of current measured by the (a-c) AHE and (d-f) ANE of SiO$_2$/Mn$_3$Sn(50nm)/Pt(5nm)/Si$_3$N$_4$(5nm) (s85) at different in-plane fields. (g) AHE switching amplitude $\Delta R_{\rm xy}^{\rm 1f}$ measured at $\pm24$~V (circles) and ANE amplitude $R_{\rm xy}^{\rm 2f}$ vs in-plane field. 
			}
		\end{figure*}
		
		We performed switching measurements while simultaneously monitoring the AHE and ANE after pulsing the sample with 1-$\mu$s-long pulses of different amplitude and $\tau \approx 130$~ns, as shown in Fig.~\ref{fig:11}. 
		For very low in-plane field $B_{\rm x}=-3$~mT the sample is switched to the intermediate state with average $|\mathbf{g}|\approx0$, as both the AHE [Fig.~\ref{fig:11}~(a)] and ANE [Fig.~\ref{fig:11}~(d)] reach the intermediate level. On the other hand, when pulsing in the presence of strong positive or negative $B_{\rm x}$, the \textbf{g}-vector switches to states with a well-defined $g_{\rm z}$ component determined by the relative alignment of current and $B_{\rm x}$ [Fig.~\ref{fig:11}~(b,c)], as expected for spin-orbit torque switching, and also a net $g_{\rm x}$ component parallel to $B_{\rm x}$ [Fig.~\ref{fig:11}~(e,f)]. The $g_{\rm x}$ component cannot be reversibly switched by the current and saturates for $B_{\rm x} \geq 200$~mT. The AHE switching amplitude reduces once the ANE saturates [Fig.~\ref{fig:11}~(g)], indicating that the reduction of the $g_{\rm z}$ component beyond 200~mT discussed in the main text is related to the emergence of a significant fraction of domains with in-plane orientation of the \textbf{g}-vector induced by $B_{\rm x}$.
		
	\section{SM7. Dependence of the switching amplitude and offset on the direction of the magnetic field}
	\begin{figure*}[h]
		\includegraphics[width=\textwidth,keepaspectratio]{SM_Fig12_tilt_dependence_3.eps}
		\label{fig:12}
		\caption{Switching loops of SiO$_2$/Mn$_3$Sn(50nm)/Pt(5nm)/Si$_3$N$_4$(5nm) (s85) in a tilted field. The pulses are 10~$\mu$s long with rise/fall time of (a) $\tau = 420$~ns and (b) $\tau = 35$~ns. The angle $\theta$ is indicated in the legend. }
	\end{figure*}

	To achieve SOT switching, the magnetic field required to break the symmetry between $+z$ and $-z$ states is usually applied along the \textbf{x} axis, parallel to the current, similar to ferromagnets with perpendicular magnetic anisotropy \cite{Miron2011,Manchon2019}. However, as noted in the main text, the switching amplitude $\Delta R_{\rm xy}$ measured at $\pm 2$~V has a nonmonotonic behavior, increasing sharply up to $B_{\rm x} \approx 100$~mT and significantly decreasing as $B_{\rm x} \gtrsim 400$~mT. This behavior is consistent with SOT switching of the \textbf{g}-vector between $+z$ and $-z$ states being promoted by a moderate $B_{\rm x}$, and the in-plane reorientation of the \textbf{g}-vector along $B_{\rm x}$ at strong fields (see Sect. SM4), which also happens during cool down at the end of a current pulse. 
	
	Additionally, if the applied magnetic field has a nonzero out-of-plane component $B_{\rm z}$, the fraction of grains with \textbf{g}-vector pointing parallel to $B_{\rm z}$ increases after a current pulse, offsetting the entire current-switching loop of $R_{\rm xy}$. In order to prove this point, we have measured the switching loops as a function of current for positive and negative external fields applied at an angle $\theta$ relative to the sample normal. Figure~\ref{fig:12} shows that the SOT switching amplitude is largest when the field is perfectly in-plane ($\theta=90^{\circ}$) and reduces as the field tilts out-of-plane. Moreover, there is also a significant shift in the AHE offset due to the nonzero $B_{\rm z}$ component re-orienting some grains during the cooldown from $T_{\rm N}$. This effect is more prominent for the short pulse fall scenario, as expected. The offset may lead to a change of sign of $\Delta R_{\rm xy}$ measured at $\pm 20$~V relative to $\Delta R_{\rm xy}$ measured at $\pm 2$~V [cfr. Fig. 3(d) in the main text].


		\bibliography{Mn3Sn_paper_2}